\documentclass[%
 reprint,
superscriptaddress,
 amsmath,amssymb,
 aps,
 longbibliography,
prb
]{revtex4-2}

\usepackage{graphicx}
\usepackage{dcolumn}
\usepackage{bm}
\usepackage{xcolor}
\usepackage{times}
\usepackage[normalem]{ulem}
\usepackage{units}
\usepackage[hypertexnames=false,linktocpage=true,colorlinks=true,linkcolor=blue,anchorcolor=blue,citecolor=blue,filecolor=blue,urlcolor=blue,bookmarksnumbered=true,pdfview=FitB,breaklinks=true]{hyperref}

\begin{document}
\title{Low temperature magnetic structure and lattice response in SmCuAs$_2$}

\author{M.~G.~Kim}\email{mgkim@uwm.edu}
\affiliation{Department of Physics, University of Wisconsin-Milwaukee, Milwaukee, WI 53201, USA}

\author{C.~Neupane}
\affiliation{Department of Physics, University of Wisconsin-Milwaukee, Milwaukee, WI 53201, USA}

\author{Y. Yu}
\affiliation{Department of Physics, University of Wisconsin-Milwaukee, Milwaukee, WI 53201, USA}

\author{R. Acevedo-Esteves}
\affiliation{National Synchrotron Light Source II, Brookhaven National Laboratory, Upton, New York 11973, USA}

\author{C. Nelson}
\affiliation{National Synchrotron Light Source II, Brookhaven National Laboratory, Upton, New York 11973, USA}

\author{D. Evans}
\affiliation{Department of Physics, Simon Fraser University, Burnaby, British Columbia, Canada}

\author{E.~D.~Mun}
\affiliation{Department of Physics, Simon Fraser University, Burnaby, British Columbia, Canada}

\author{D. F. Agterberg}
\affiliation{Department of Physics, University of Wisconsin-Milwaukee, Milwaukee, WI 53201, USA}

\author{J.-W. Kim}
\affiliation{Advanced Photon Source, Argonne National Laboratory, Argonne, Illinois 60439, USA}

\date{\today}


\begin{abstract}

We investigated the structural and magnetic properties of single-crystalline SmCuAs$_2$ using high-resolution synchrotron X-ray diffraction and X-ray resonant magnetic scattering (XRMS) at the Sm $L_2$ and $L_3$ edges. Temperature-dependent diffraction measurements confirm that SmCuAs$_2$ maintains its tetragonal symmetry from room temperature down to 8 K, with lattice parameters showing anomalous behavior below the resistivity minimum ($T \approx$ 30 K). Notably, the \textbf{c}-axis lattice parameter exhibits a plateau and subsequent increase near the N\'{e}el temperature, indicating magnetoelastic coupling. XRMS measurements reveal a commensurate antiferromagnetic structure with a propagation vector \textbf{\textit{q}} = (0, 0, 0.5). Our measurement shows that the Sm moments are aligned within the \textbf{\textit{ab}} plane and arranged in a $++--$ stacking along the \textbf{\textit{c}}-axis. Comparison with related \textit{RE}CuAs$_2$ compounds (\textit{RE} = Pr, Nd, and Gd) suggests that in-plane moment orientation and associated magnetic frustration play a key role in the emergence of the resistivity minimum. Differences in spin-orbit and magnetoelastic coupling across the series highlight their importance in governing low-temperature transport behavior.

\end{abstract}

\maketitle

The family of \textit{RE}CuAs$_2$ (\textit{RE} = rare-earth) shows interesting and complex physics.  \textit{RE}CuAs$_2$ with \textit{RE} = Nd, Sm, Gd, Tb, and Dy, which are not Kondo materials, produce resistivity minimum at low temperature before the appearance of antiferromagnetic ordering.\cite{Sampathkumaran-2003,SENGUPTA2004465, Evans} The Kondo effect is generally expected to be strongly suppressed in systems with large local magnetic moments or with strongly easy-axis spin anisotropy as in \textit{RE}CuAs$_2$ with \textit{RE} = Nd, Sm, Gd, Tb, and Dy. Thus, the observed resistivity minimum in these materials cannot be explained by the Kondo effect through an exchange interaction between isolated \textit{f}-electrons and conduction \textit{d}-electrons.~\cite{MAPLE-1}  It was recently proposed that strong frustration stabilizing liquid-like spin states can induce a resistivity minimum in non-Kondo systems.~\cite{Wang-2016} This state is characterized by an enhanced spin structure factor at wave vectors smaller than twice the Fermi wave vector. Several alternative explanations for the resistivity minimum stem from the spin-slip scattering due to various mechanisms related to the underlying magnetic properties~\cite{bulk-ceramic-1, gran-2,Das-2015,spin-2, Matsushita-2005, polaron1,spin-3,polaron2,polaron3,spin-4,spin-5,spin-6,spin-7}

Understanding the resistivity minimum in this family of compounds necessitates knowledge of both the crystal and magnetic structures and their behaviors at low temperatures. Previous research on polycrystalline samples revealed the tetragonal $P4/nmm$ structure of \textit{RE}CuAs$_2$ at room temperature\cite{Mozharivskyj-2000,Mozharivskyj-2002,JEMETIO200293,Sampathkumaran-2003,SENGUPTA2004465}. The room temperature crystal structure of GdCuAs$_2$ has been studied to change from the tetragonal structure to an orthorhombic $Pmmn$ structure with slight P doping. However, the crystal structures of this family have not been investigated at low temperatures. Recent investigation on the low temperature crystal structure of GdCuAs$_2$ reveals the orthorhombic structure.~\cite{Ashiwini}

The magnetic structure of this family was investigated using neutron powder diffraction (NPD).\cite{Zhao-2017} It reveals that the antiferromagnetic (AFM) structure of PrCuAs$_2$ (no resistivity minimum) with the Pr moments pointing along the \textit{\textbf{c}} axis with \textbf{\textit{q}} = (0,~0,~0.5). In NdCuAs$_2$ and DyCuAs$_2$, both of which exhibit resistivity minimum,\cite{Sampathkumaran-2003,SENGUPTA2004465, Evans} their magnetic moments lie in the \textit{\textbf{ab}} plane with \textit{\textbf{q}} = (0,~0,~0.5).\cite{Zhao-2017} For TbCuAs$_2$ (exhibiting resistivity minimum) and HoCuAs$_2$ (resistivity minimum unknown), the study finds a complex incommensurate magnetic ordering with propagation vectors \textbf{\textit{q$_1$}}(Tb) = (0.240,~0.155,~0.48) and \textbf{\textit{q$_2$}}(Tb) = (0.205,~0.115,~0.28) and \textbf{\textit{q}}(Ho) = (0.121,~0.041,~0.376), respectively, without a specific model of their magnetic structures. Due to the intrinsic limitation of the NPD, the details of the magnetic structures of those materials have not been determined, such as the moment arrangement along the \textit{\textbf{c}} axis in \textit{RE}CuAs$_2$ (\textit{RE} = Pr, Nd, and Dy). Recent single crystal study on GdCuAs$_2$ by the X-ray resonant magnetic scattering (XRMS) shows that GdCuAs$_2$ orders antiferromagnetically below $T_\mathrm{N_1} \approx$ 10 K with incommensurate propagation vector \textbf{\textit{q}} = ($\delta$, 0, 0.5) followed by a lock-in transition at $T_\mathrm{N_2} \approx$ 6 K to a commensurate AFM ordering at \textbf{\textit{q}} = (1/3, 0, 0.5). The magnetic structure obtained for GdCuAs$_2$ differs from those proposed for its sister compounds by NPD studies.~\cite{Ashiwini,Zhao-2017}

In this paper, we study the low temperature magnetic structure and lattice response in SmCuAs$_2$, which exhibits the resistivity minimum at around 30 K. We perform high-resolution X-ray diffraction measurements to investigate the lattice parameters as a function of temperatures. We find that the lattice parameters \textit{a} and \textit{c} decrease linearly as temperature lowers and reach their minimum or plateau near the temperature where the resistivity minimum occurs. We also observed an increase in the $c$ lattice parameter below the N\'{e}el temperature, indicating the presence of strong magnetoelastic coupling. We also utilize the XRMS to determine the precise magnetic moment orientation and its arrangement. We observed that the magnetic Bragg peak appears at (0, 0, 0.5) below 12 K, and polarization-dependent measurements revealed that the Sm moments are aligned in the \textbf{\textit{ab}}-plane, forming an antiferromagnetic coupling of the $++--$ type along the \textbf{\textit{c}}-axis, similar to the structure observed in NdCuAs$_2$ and DyCuAs$_2$. Comparison between SmCuAs$_2$ and GdCuAs$_2$ indicate a strong interplay between resistivity anomalies, lattice distortion, and magnetic structure.

Single crystals of SmCuAs$_2$ have grown from high-temperature ternary melts.~\cite{Evans, Ashiwini} The phase purity of the samples from the growth batch was examined by X-ray powder diffraction using a Rigaku Miniflex at room temperature. Magnetic and electrical properties of as-grown single crystals were checked using Quantum Design Magnetic Property Measurement System and a Quantum Design Physical Properties Measurement System.  

Temperature-dependent XRMS measurements and high-resolution single-crystal X-ray diffraction measurements were performed on a six-circle diffractometer at the integrated in situ and resonant hard x-ray studies (4-ID) beam line of National Synchrotron Light Source II (NSLS-II), which allows convenient access to a wide region of reciprocal space. Measurements were performed at the Sm $L_2$ ($E$~=~7.316~keV) and $L_3$ edge ($E$~=~6.712~keV). An as-grown plate-like single crystal was attached to a flat copper sample holder on the cold finger of a closed cycle refrigerator (the base temperature $T \approx$ 8 K) and initially mounted to be the [1, 0, 0] - [0, 0, 1] within the scattering plane.  The mosaicity of the SmCuAs$_2$ single crystal was less than 0.04$^\circ$ full-width-at-half-maximum as measured by the rocking curve of the (0, 0, 6) reflection at room temperature. The diffraction data were obtained as a function of temperature between room temperature and $T \approx$ 8 K, the base temperature of the refrigerator. For XRMS, by utilizing the six-circle diffractometer, we can rotate the sample azimuth angle to constrain the [0, 0, 1] in the scattering plane. The incident radiation was linearly polarized perpendicular to the vertical scattering plane ($\sigma$ polarized). We utilized a Pyrolytic Graphite analyzer for polarization analysis. For example. in this configuration, dipole resonant magnetic scattering at Sm $L_2$ rotates the scattered beam polarization into the scattering plane ($\pi$ polarization), and the magnetic moment direction can be precisely determined. The magnetic moment components in the scattering plane (moment components projected onto the scattering plane) contribute to scattering intensity, while the magnetic moment components perpendicular to the scattering plane produce zero intensity. Scattered intensities were recorded using a two-dimensional detector.

\begin{figure}[!t]
    \centering
    \includegraphics[width=1\linewidth]{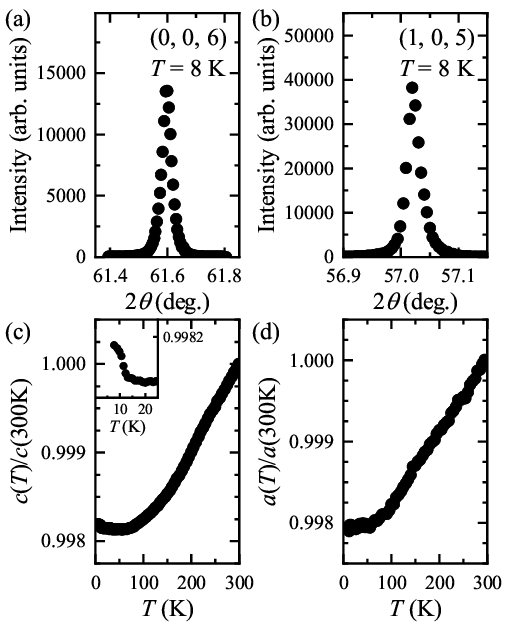}\\
    \caption{(a) Longitudinal scans ($\theta - 2\theta$ scans) of the (0, 0, 6) Bragg peak measured at 8 K. (b) Rocking scan at the (1, 0, 5) at $T =$ 8 K. (c) Lattice parameter $c$ extracted from (0, 0, 6) Bragg peak as a function of temperature. The inset shows an expanded view of the data below 25 K. (d) Lattice parameter $a$ calculated from the (1, 0, 5) Bragg peak as a function of temperature.} 
    \label{fig1}
\end{figure}

Figure~\ref{fig1} shows our high-resolution single-crystal X-ray diffraction results. Figures~\ref{fig1} (a) and (b) show longitudinal scans ($\theta - 2\theta$ scans) of the charge peaks at (0, 0, 6) and (1, 0, 5), respectively. Both peaks are observed to be very sharp, and their sharpness remains unchanged with temperature, indicating the absence of structural transitions. From such temperature-dependent measurements, the lattice parameters $c$ and $a$ can be extracted. The resulting lattice parameters are presented in Figs.~\ref{fig1} (c) and (d). Both the lattice parameters $a$ and $c$ decrease linearly with decreasing temperature until they approach the temperature ($T \approx$ 30 K) at which the resistivity minimum occurs, where the rate of decrease slows down. Notably, the lattice parameter $c$ appears to plateau or even slightly increase below the resistivity minimum temperature. At the N\'{e}el temperature, where AFM order sets in, the lattice parameter $a$ shows little to no change, whereas $c$ exhibits an increase. This behavior is reminiscent of magnetoelastic coupling observed in GdCuAs$_2$.~\cite{Ashiwini} However, a key difference is that in SmCuAs$_2$, the lattice parameter $a$ remains unchanged, indicating the absence of a structural transition, in contrast to what is observed in GdCuAs$_2$. It is noteworthy that we cannot entirely rule out other types of structural distortion, such as local symmetry breaking. Nevertheless, this contrast likely originates from differences in the magnetic structure, which will be discussed next.

\begin{figure}[!t]
    \centering
    \includegraphics[width=1\linewidth]{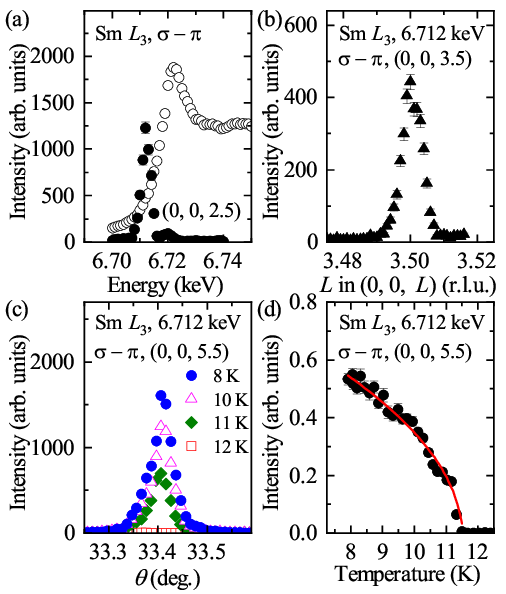}\\
    \caption{(a) Fluorescence (open symbols) and energy scan (closed symbols) measured at the Sm $L_3$ edge. The energy scan was performed at the antiferromagnetic (AFM) Bragg peak (0, 0, 2.5) in the $\sigma - \pi$ channel. (b) L-scan of the (0, 0, 3.5) reflection measured at the Sm $L_3$ edge. (c) Rocking scans of the AFM Bragg peak (0, 0, 5.5) measured at several representative temperatures. (d) AFM order parameter extracted from the (0, 0, 5.5) AFM Bragg peaks at the Sm $L_3$ edge.}
    \label{fig2}
\end{figure}

The XRMS results measured at the Sm $L_3$ absorption edge are shown in Fig.~\ref{fig2}. Figure~\ref{fig2} (a) presents both the fluorescence spectrum and the energy scan performed at the Sm $L_3$ edge. The energy scan was measured at the (0, 0, 2.5) magnetic reflection in the $\sigma - \pi$ geometry and reveals a characteristic double-peak feature. The first, more intense peak appears at $E$ = 6.712 keV, while a weaker second peak emerges at $E$ = 6.72 keV. This double-peak structure is commonly observed in XRMS measurements at the Sm $L_3$ edge. The lower-energy peak is attributed to the electric quadrupole ($E$2) $2p - 4f$ transition, which typically appears a few eV below the absorption edge due to the stronger core-hole–excited electron interaction in the intermediate state. The higher-energy peak corresponds to the electric dipole ($E$1) transition from the $2p$ core level to the unoccupied $5d$ states.~\cite{reson-1,reson-2,reson-3}

Figures~\ref{fig2} (b) and (c) display magnetic Bragg peaks at (0, 0, 3.5) and (0, 0, 5.5), respectively. The $L$-scan through (0, 0, 3.5) shows that the AFM Bragg peaks appear at half-integer values of $L$. In contrast to GdCuAs$_2$, no AFM peaks were observed along the $H$ or $K$ directions, clearly indicating that the propagation vector is \textbf{\textit{q}} = (0, 0, 0.5). To investigate the temperature dependence of the magnetic signal, we measured the (0, 0, 5.5) peak as a function of temperature, shown in Fig.~\ref{fig2} (c), and extracted the AFM order parameter, shown in Fig.~\ref{fig2} (d). As the temperature decreases, the AFM Bragg peak at (0, 0, 5.5) emerges below $T_N =$ 11.5(5) K and grows in intensity. Since the order parameter does not appear to saturate down to our base temperature of 8 K, the magnetic moment likely continues to increase below this temperature. The slightly lower $T_N$ observed in our XRMS measurement compared to the value of $T_N \approx$ 12 K obtained from bulk measurements~\cite{Sampathkumaran-2003,SENGUPTA2004465, Evans} is likely due to differences in the temperature sensors used in the two setups.

\begin{figure}[!t]
    \centering
    \includegraphics[width=1\linewidth]{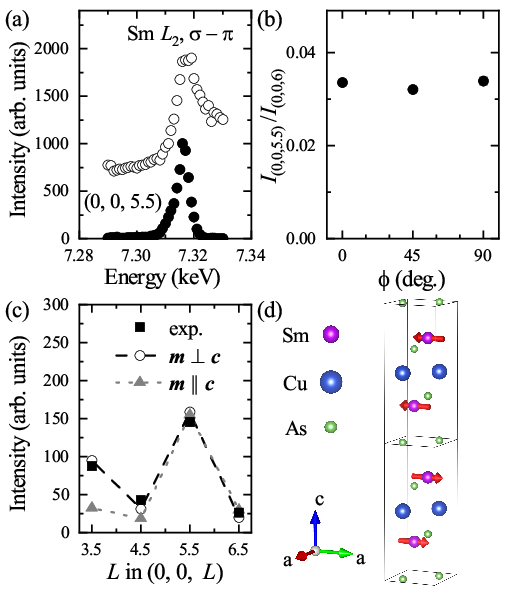}\\
    \caption{(a) Fluorescence (open symbols) and energy scan (closed symbols) measured at the Sm $L_2$ edge. The energy scan was performed at the AFM (0, 0, 5.5) Bragg peak in the $\sigma - \pi$ channel. (b) Azimuthal dependence of the intensity ratio between the AFM (0, 0, 5.5) Bragg peak and the structural (0, 0, 6) peak. (c) $L$-dependence of the AFM (0, 0, $L$) Bragg peaks. Squares represent experimental data, open circles show calculated intensities for $\textbf{\textit{m}}\perp\textbf{\textit{c}}$ ($\Gamma_{10}$), and triangles represent calculated intensities for $\textbf{\textit{m}}~\|~\textbf{\textit{c}}$ ($\Gamma_{3}$). (d) Schematic illustration of the AFM structure of SmCuAs$_2$.}
    \label{fig3}
\end{figure}

The electric dipole ($E$1) transition is dominant at the $L_2$ resonance, and in the case of commensurate AFM order, only the odd harmonics contribute to the magnetic scattering.~\cite{reson-1,reson-2,reson-3} This allows for a simplified analysis in which only the first-order term in the resonant matrix element needs to be considered.~\cite{Blume,Hannon,Hill} Taking advantage of this, we conducted a detailed study of the magnetic structure at the Sm $L_2$ edge. Figure~\ref{fig3} (a) presents the fluorescence scan measured at the Sm $L_2$ edge along with the energy scan at the AFM Bragg peak (0, 0, 5.5) in the $\sigma - \pi$ geometry. A clear $E$1 resonance is observed at $E$ = 7.316 keV. This is consistent with the AFM magnetic order observed at Sm $L_3$ edge in Fig.~\ref{fig2}. 

To determine the Sm magnetic moment arrangement more precisely, we performed azimuthal scans and measured the \textbf{\textit{q}}-dependence of the AFM Bragg peak. These measurements were conducted at the AFM Bragg reflection (0, 0, 5.5). For the azimuthal dependence, the intensity of the charge peak at (0, 0, 6) was also measured to calculate the intensity ratio, thereby eliminating the influence of domain population changes caused by variations in the illuminated area during sample rotation. The azimuthal dependence is shown in Figure~\ref{fig3} (b). We observed that the AFM magnetic intensity remains constant as the sample is rotated from 0$^\circ$ to 90$^\circ$. The absence of azimuthal dependence in the Sm magnetic moment signal may be interpreted as evidence that the Sm moments are aligned along the \textbf{c}-axis with no in-plane component. However, as previously discussed, SmCuAs$_2$ retains a tetragonal structure at all measured temperatures, and thus the result may also be attributed to the presence of magnetic domains with an in-plane Sm moment component. Specifically, if the Sm moments possess an in-plane component and the sample is in a single-domain state, the azimuthal dependence would follow a sinusoidal variation ($I \propto \cos^2\phi$). However, in the presence of multiple magnetic domains, such variations would average out, resulting in a flat azimuthal response as observed in Fig.~\ref{fig3} (b).

We measured the \textbf{\textit{q}}-dependence of the AFM (0, 0, $L$) Bragg peaks to determine the Sm moment arrangement, as shown in Fig.~\ref{fig3} (c). The possible magnetic structures were calculated using representation analysis.~\cite{WILLS2000} For the space group $P4/nmm$ with propagation vector \textbf{\textit{q}} = (0, 0, 0.5), there are $\Gamma_2, \Gamma_3, \Gamma_9$ and $\Gamma_{10}$ representations (MRs), consistent with Ref.~\cite{Zhao-2017}. $\Gamma_2$ and $\Gamma_3$ correspond to Sm moments aligned along the \textbf{\textit{c}}-axis, whereas $\Gamma_9$ and $\Gamma_{10}$ describe moments oriented in the in-plane direction. Another key distinction among these MRs lies in the arrangement of Sm moments along the \textbf{\textit{c}}-axis: in $\Gamma_2$ and $\Gamma_9$, the moments follow a $+--+$ configuration, while in $\Gamma_3$ and $\Gamma_{10}$, they adopt a $++--$ stacking sequence. We calculated the AFM (0, 0, $L$) Bragg peak intensities for the MRs using the relation $I \propto |F|^2$, where the structure factor $F$ was obtained as $F \propto \sum f^{XRMS}e^{i \mathbf{k} \cdot \mathbf{r}}$. Here, $f^{XRMS} = f^{XRMS}_{\sigma - \pi} \propto m_1\cos\theta + m_3\sin\theta$ with $m_1$ and $m_3$ representing the magnetic moment (\textbf{\textit{m}}) components along the \textbf{\textit{a}} and \textbf{\textit{c}} axes, respectively; \textbf{\textit{k}} is the reciprocal lattice vector, and \textbf{\textit{r}} is the real space positions of magnetic elements. In these calculations, for representations such as $\Gamma_2$ or $\Gamma_9$ with a $+--+$ stacking, the Bragg peaks at $L$ = (even + 0.5) are more intense than those at $L$ = (odd + 0.5). In contrast, for $\Gamma_3$ or $\Gamma_{10}$ with a $++--$ stacking, the $L$ = (odd + 0.5) peaks are stronger than the $L$ = (even + 0.5) ones, consistent with our experimental observations. Therefore, $\Gamma_2$ and $\Gamma_9$ can be excluded as possible magnetic structures.

The calculated $L$-dependence of the AFM Bragg peak intensity for $\Gamma_3$ (Sm moments along the \textbf{\textit{c}}-axis, $\textbf{\textit{m}}~\|~\textbf{\textit{c}}$) and $\Gamma_{10}$ (Sm moments in the \textbf{\textit{ab}}-plane, $\textbf{\textit{m}}\perp\textbf{\textit{c}}$), incorporating the moment direction, is shown in Fig.\ref{fig3} (c). As illustrated, the experimental data agree closely with the calculation for in-plane moment alignment ($\textbf{\textit{m}}\perp\textbf{\textit{c}}$), which is $\Gamma_{10}$ MR. However, due to the influence of magnetic domains as discussed earlier, the precise in-plane orientation of the Sm moments ($\textbf{\textit{m}}\perp\textbf{\textit{c}}$) within the $\Gamma_{10}$ representation cannot be determined from azimuthal dependence. Taken together, SmCuAs$_2$ adopts a magnetic structure described by the $\Gamma_{10}$ representation, in which Sm moments are aligned within the \textbf{\textit{ab}}-plane ($\textbf{\textit{m}}\perp\textbf{\textit{c}}$) and are arranged along the \textbf{\textit{c}}-axis in a $++--$ sequence. The resulting magnetic structure is depicted in Fig.~\ref{fig3} (d). Note that the magnetic susceptibility measured at $H$ = 1~kOe~\cite{Evans} suggests that the moments may be aligned along the \textbf{\textit{c}} axis.

The AFM ordering wave vector of SmCuAs$_2$ is \textbf{\textit{q}} = (0, 0, 0.5), a commensurate values similar to those observed in PrCuAs$_2$ and NdCuAs$_2$. While PrCuAs$_2$ does not exhibit a low-temperature resistivity minimum, both NdCuAs$_2$ and SmCuAs$_2$ do.\cite{Sampathkumaran-2003,SENGUPTA2004465, Evans} A key distinction among these compounds lies in the direction of their magnetic moments: PrCuAs$_2$ has moments aligned along the \textbf{\textit{c}}-axis ($\textbf{\textit{m}}~\|~\textbf{\textit{c}}$),\cite{Zhao-2017} whereas both NdCuAs$_2$~\cite{Zhao-2017} and SmCuAs$_2$ possess in-plane moments ($\textbf{\textit{m}}\perp\textbf{\textit{c}}$). This suggests that the direction of the magnetic moments and the resulting magnetic anisotropy may be a crucial factor. Further support for this idea is provided by GdCuAs$_2$, which exhibits an incommensurate AFM ordering with a wave vector of \textbf{\textit{q}} = ($\delta$, 0, 0.5) and in-plane Gd moments ($\textbf{\textit{m}}\perp\textbf{\textit{c}}$), and also displays a low-temperature resistivity minimum.\cite{Ashiwini} 

In our study of SmCuAs$_2$, the in-plane moments ($\textbf{\textit{m}}\perp\textbf{\textit{c}}$) are frustrated due to the degeneracy of their orientations arising from the four-fold symmetry of the tetragonal lattice. While such a frustration is relieved in GdCuAs$_2$ by forming orthorhombic structure, the incommensurate nature of its magnetic ordering suggests that significant magnetic frustration remains in GdCuAs$_2$ despite the structural distortion. Such frustration in both types of AFM structures is consistent with theoretical models suggesting that resistivity minima in frustrated itinerant antiferromagnets may originate from enhanced spin scattering mediated by the RKKY interaction.\cite{Wang-2016} However, we cannot assess how the spin fluctuations may affect the electronic states in this family of compounds and further studies such as inelatic neutron scattering measurements are called for. These findings suggest that magnetic frustration may play a critical role in the emergence of the resistivity minimum in the \textit{RE}CuAs$_2$ family.

Our previous observations on GdCuAs$_2$ reveal an unusually strong magnetoelastic coupling.\cite{Ashiwini} This is particularly surprising given that Gd$^{3+}$ ions are generally considered to possess very weak spin-orbit coupling, and consequently, minimal coupling to the lattice\cite{rareearth1, rareearth2}, although spin-lattice coupling has been reported in some Gd-based compounds\cite{ref01,ref02,ref03}. Observed strong magnetoelastic effects are therefore not anticipated. In contrast, SmCuAs$_2$ is expected to exhibit fairly strong spin-orbit coupling and, correspondingly, appreciable coupling to the lattice.\cite{rareearth1, rareearth2} However, our experimental results show no evidence of a structural transition, indicating that magnetoelastic coupling in SmCuAs$_2$ is quite weak. This may imply that the effective spin-orbit coupling in SmCuAs$_2$ is significantly smaller than anticipated, which could explain the observed insensitivity of its resistivity to magnetic fields up to 9 T.\cite{Evans} Conversely, the fact that the resistivity minimum in GdCuAs$_2$ is suppressed even by relatively small magnetic fields\cite{Ashiwini} suggests the possibility of unexpectedly strong spin-orbit effects in this compound. These contrasting behaviors underscore the critical role of spin-orbit and magnetoelastic coupling in the \textit{RE}CuAs$_2$ family and point to a strong interplay between resistivity anomalies, lattice distortion, and moment direction. Further experimental and theoretical studies, such as those involving thermal expansion, magnetostriction, or the role of spin-orbit coupling, may shed light on the underlying mechanism.

In summary, we investigated the magnetic and structural properties of SmCuAs$_2$ single crystals using high-resolution X-ray diffraction and X-ray resonant magnetic scattering. Temperature-dependent diffraction data revealed no structural transition, but showed anomalous behavior in the lattice parameter $c$ near the resistivity minimum ($T \approx$ 30 K) and N\'{e}el temperature ($T_N =$ 11.5(5) K), indicating magnetoelastic coupling. XRMS measurements at the Sm $L_2$ and $L_3$ edges identified commensurate antiferromagnetic order with a propagation vector \textbf{\textit{q}} = (0, 0, 0.5). We determined the magnetic structure of SmCuAs$_2$, revealing in-plane moments ($\textbf{\textit{m}}\perp\textbf{\textit{c}}$) arranged in a $++--$ stacking sequence along the \textbf{\textit{c}}-axis. By comparing this to other \textit{RE}CuAs$_2$ compounds (\textit{RE} = Pr, Nd, and Gd), we highlight the crucial role of the in-plane moment alignment and the magnetic frustration in the emergence of a low-temperature resistivity minimum. Furthermore, by examining the structural distortions and the response of resistivity to an external magnetic field, we discuss the importance of spin-orbit and magnetoelastic coupling in this system.\cite{data}

\begin{acknowledgments}
This work was supported by the University of Wisconsin-Milwaukee. 
This work used resources at the 4-ID beamline of the National Synchrotron Light Source II, a U.S. Department of Energy (DOE) Office of Science User Facility operated for the DOE Office of Science by Brookhaven National Laboratory under Contract No. DE-SC0012704. 
Y. Y. and D.F.A. were supported by the National Science Foundation Grant No. DMREF 2323857. 
This research used resources from the Advanced Photon Source, a U.S. Department of Energy (DOE) Office of Science User Facility operated for the DOE Office of Science by Argonne National Laboratory under Contract No. DE-AC02-06CH11357.
E. Mun was supported by the Canada Research Chairs, Natural Sciences and Engineering Research Council of Canada, and Canada Foundation for Innovation program.

\end{acknowledgments}

\bibliographystyle{apsrev4-2-title}
\bibliography{SmCuAs2}

\end{document}